\documentclass[prb,showpacs,twocolumn]{revtex4}
\usepackage{tabularx}
\usepackage{amsmath}
\usepackage{amssymb}
\usepackage{graphicx}
\usepackage{dcolumn}
\usepackage{bm}

\begin{document}

\title{Signature of Majorana Fermions in Charge Transport in Semiconductor
Nanowires}
\author{Chunlei Qu$^{1}$}
\author{Yongping Zhang$^{1}$$^,$$^{2}$}
\author{Li Mao$^{1}$}
\author{Chuanwei Zhang$^{1}$}
\thanks{Author to whom correspondence should be addressed: cwzhang@wsu.edu}
\affiliation{$^{1}$Department of Physics and Astronomy, Washington State University,
Pullman, WA, 99164 USA \\
$^{2}$ICQS, Institute of Physics, Chinese Academy of Sciences, Beijing
100190, China}

\begin{abstract}
We investigate the charge transport in a semiconductor nanowire that is
subject to a perpendicular magnetic field and in partial contact with an
\textit{s}-wave superconductor. We find that Majorana fermions, existing at
the interface between superconducting and normal sections of the nanowire
within certain parameter region, can induce resonant Andreev reflection of
electrons at the interface, which yields a zero energy peak in the
electrical conductance of the nanowire. The width of the zero energy
conductance peak for different experimental parameters is characterized.
While the zero energy peak provides a signature for Majorana fermions in one
dimensional nanowires, it disappears in a two-dimensional semiconductor thin
film with the same experimental setup because of the existence of other edge
states in two dimensions. The proposed charge transport experiment may
provide a simple and experimentally feasible method for the detection of
Majorana fermions in semiconductor nanowires.
\end{abstract}

\maketitle

% 74.45.+c      proximity effects; Andreev reflection; SN and SNS junctions
% 73.63. Nm     Quantum wires
%03.67.Lx       Quantum computation architectures and implementations

\section{Introduction}

Majorana fermions are quantum particles which are their own anti-particles%
\cite{Majorana}, unlike ordinary Dirac fermions where electrons and
positrons (holes) are distinct. Recently, some exotic ordered states in
condensed matter systems such as the Pfaffian states in fractional quantum
Hall (FQH) systems\cite{FQH1,FQH2,FQH3,FQH4}, chiral \textit{p}-wave
superconductors such as strontium ruthenate\cite{Ivanov,SR,Stone}, chiral
\textit{p}-wave fermionic superfluids from \textit{p}-wave Feshbach resonance%
\cite{Melo,Gurarie,Yip,Tewari2007} , surface of a 3D strong topological
insulator (TI){\cite{Fu1,Fu2,Akhmerov,Nagaosa1}}, as well as
semiconductor/superconductor heterostructures have been proposed as
platforms supporting Majorana fermions\cite%
{Sau1,Sau2,Alicea,Roman1,Oreg,Roman2,Mao1,Mao2,Mao3}. The commonality
between these systems is that they all allow quasiparticle excitations which
involve no energy cost (the quasiparticle energy is exactly at the Fermi
energy). The second quantized operators, $\gamma _{i}$, of these zero-energy
excitations are self-hermitian, $\gamma _{i}^{\dagger }=\gamma _{i}$, which
is the defining property of Majorana fermions. Essentially this
self-hermitian property of Majorana fermions leads to a very non-trivial
exchange statistics of these particles called non-Abelian statistics\cite%
{kitaev,Nayak}. Although the emergence of Majorana excitations in
low-temperature physical systems would by itself be a truly extraordinary
phenomenon, they have also sparked tremendous recent interest because of
their potential use in fault tolerant topological quantum computation (TQC)%
\cite{kitaev,Nayak}.

Despite the theoretical success, Majorana fermions have proven to be hard to
observe in natural physical systems, such as chiral \textit{p}-wave
superconductors/superfluids\cite{SR,Tewari2007}. To circumvent this problem,
it has been proposed that Majorana fermions may be realized in much more
physically robust \textit{s}-wave superconductors/superfluids by utilizing
two additional components: Rashba spin-orbit coupling and Zeeman fields\cite%
{Zhang,Sato}. In this context, it has been shown that a semiconducting thin
film or nanowire with a sizable spin-orbit coupling and in proximity contact
with an \textit{s}-wave superconductor, can host Majorana fermion
excitations localized near defects in the presence of a suitable Zeeman
splitting \cite{Sau1,Sau2,Alicea,Roman1,Oreg,Roman2,Mao1,Mao2,Mao3}.

With the theoretical validation of the existence of Majorana fermions in
such semiconductor/superconductor heterostructures, one natural and
important question to ask is how to detect Majorana fermions in experiments.
Currently two types of experimental schemes have been proposed: the
tunneling conductance (e.g., scanning tunneling microscope (STM)) \cite%
{Sau2,Mao1,Mao3,Flensberg} and the Josephson junction\cite{Roman1,Oreg}. It
was shown that there is a zero energy peak in the STM tunneling conductance
when the underlying heterostructure is in a topological state with Majorana
fermions, while the peak disappears when the heterostructure is in a
non-topological state without Majorana fermions. In the Josephson junction
type of experiments, it was shown that the Josephson current has a period of
4$\pi $ in the presence of Majorana fermions, in contrast to the 2$\pi $
period for a regular \textit{s}-wave superconductor without Majorana fermions%
\cite{Roman1,Kitaev2,Fu3,Kwon}. However, these experiments are generally
very sensitive to temperature and external noise and are hard to realize.
Nowadays, the observation of Majorana fermions is not yet reported
experimentally.

In this paper, we propose to detect Majorana fermions in a more common and
robust charge transport type of experiments \cite{Wimmer} in the
semiconductor/superconductor heterostructure. The proposed experimental
setup is illustrated in Fig. 1, where one section of the semiconductor
nanowire is in proximity contact with an \textit{s}-wave superconductor,
while the other is in contact with an insulator. The electrical conductance
is measured between the two ends of the semiconductor nanowire. At the
interface between these two sections of the nanowire, the transport of
electrons is described by the Andreev reflection (AR){\cite{Andreev}}. It is
well known that when a normal metal is in contact with a superconductor, AR
process dominates the charge transport if the energy of the incident
particle lies inside the gap of the superconductor. AR is also a useful tool
to measure spin polarization of ferromagnet \cite{Jong,Upadhyay,Soulen} and
to probe unconventional superconductor \cite{Deutscher,Sengupta,Tanaka2}.
Recently, AR has been studied in topological insulators to explore their
exotic properties \cite{Law,Linder,Tanaka,Nilsson}.

Because Majorana fermions occupy the zero energy state in the
superconductor, their signature is generally expected to be a peak of the
electrical conductance at the zero external voltage. However, in the regular
AR between a normal metal (or semiconductor) and a regular \textit{s}-wave
superconductor, the electrical conductance is already the constant $2e^{2}/h$
when the energy of the incident electron is in the superconductor gap region%
\cite{Blonder}. Therefore the signature of Majorana fermions may be hidden
in the large regular AR signals and cannot be observed in experiments. In
previous literature, the constant electrical conductance in the gap region
can be removed by adding an insulating barrier between the metal and the
superconductor\cite{Blonder}. Here we adopt the same approach and use an
insulating (or potential) barrier to prevent the regular AR between the
sections of the semiconductor nanowire with and without the contact with the
\textit{s}-wave superconductor.

In this paper, we find that when the section of the nanowire in contact with
the superconductor is in the topological state, there is a zero-bias peak in
the electrical conductance. This zero energy peak originates from the
Majorana fermions existing at the interface between the superconducting and
normal states, which induces a resonant AR of electrons although the normal
AR is strongly suppressed by the high potential barrier at the interface.
The width of the zero energy conductance peak depends on the barrier height
as well as the spin-orbit coupling strength. When the section of the
nanowire in contact with the superconductor is in the non-topological state
without Majorana fermions, there is no resonant AR at the zero energy and
the zero-bias conductance peak disappears. In this case, the electrical
conductance depends strongly on the number of conduction bands intercepted
by the chemical potential. When only one conduction band is occupied by
electrons, the electrical conductance at the zero bias is suppressed to zero
\cite{Fisher,Wimmer}, but increases with the increasing bias voltage. When
both conduction bands are occupied, the suppression of the zero bias
conductance in the single band case vanishes and the electrical conductance
is similar as that for a junction between a normal metal and a regular
\textit{s}-wave superconductor. Finally, we find that the zero-bias peak
disappears in a two-dimensional semiconductor thin film in contact with an
\textit{s}-wave superconductor where the zero energy Majorana modes coexist
with other edge modes with non-zero momentum and energy, which contribute to
the conductance signals at the non-zero energy.

The paper is organized as follows: Section II describes the proposed
experimental setup and the Hamiltonian. Section III discusses the electrical
conductance of the nanowire calculated using the Blonder-Tinkham-Klapwijk
(BTK) theory\cite{Blonder}. We study the dependence of the electrical
conductance on the spin-orbit coupling strength, the Zeeman field, and the
barrier height. We extend our model and method to a two-dimensional
semiconductor thin film in section IV. Section V consists of the discussion
and conclusion.

\section{Proposed experimental scheme and the Hamiltonian}

The proposed experimental scheme is shown in Fig. 1. The semiconductor
nanowire lying along the $x$ direction is deposited on the top of an
insulator and an $s$-wave superconductor heterostructure. Due to the
proximity effect, the $s$-wave Cooper pairs can tunnel into the right
section of the nanowire ($x>0$), yielding a non-zero superconducting pairing
order parameter $\Delta _{0}$. In the left section of the nanowire ($x<0$), $%
\Delta _{0}=0$. A magnetic field $B$ is applied along the $z$ direction,
yielding a Zeeman field $V_{z}=g_{e}^{\ast }\mu _{B}B/2$, where $g_{e}^{\ast
}$ is the Land\'{e} factor, $\mu _{B}$ is the Bohr magnet. An insulating
barrier (or potential) is applied at the interface $x=0$ between these two
sections of the nanowire. Such a potential barrier may be realized by
growing a thin junction with a different type of material around $x=0$
during the growth of the nanowire\cite{Liber,Pitanti}, or by applying a gate
voltage on the nanowire at $x=0$.

%%%%%%%%%%%%%%%%%%%%%%%%%%%%%%%%%%%%%%%%%%%%%%%%%%%%%%%%%%%%%%%%%%
\begin{figure}[t]
\includegraphics[width = 1\linewidth]{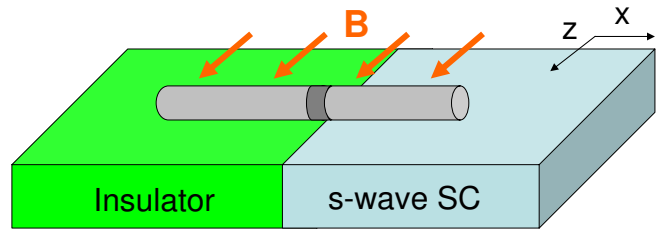}
\caption{(Color online) Illustration of the proposed experimental scheme.
The semiconductor nanowire is deposited on the top of an insulator and an
\textit{s}-wave superconductor. The dark shadow area in the nanowire
represents a potential or an insulating barrier. The magnetic field is
applied along the $z$ direction.}
\label{expsetup}
\end{figure}
%%%%%%%%%%%%%%%%%%%%%%%%%%%%%%%%%%%%%%%%%%%%%%%%%%%%%%%%%%%%%%%%%%%

The dynamics of electrons in the semiconductor nanowire can be described by
the Bogoliubov-de-Gennes (BdG) equation. In the Nambu spinor space, the BdG
equation can be written as $\mathcal{H}\Psi =E\Psi $, where the BdG
Hamiltonian

\begin{equation}
\mathcal{H}=%
\begin{pmatrix}
H_{0} & i\sigma _{y}\Delta \\
-i\sigma _{y}\Delta ^{\ast } & -H_{0}^{\ast }%
\end{pmatrix}%
,
\end{equation}%
with the quasiparticle wavefunction $\Psi =(u_{\uparrow }(x),u_{\downarrow
}(x),v_{\uparrow }(x),v_{\downarrow }(x))^{T}$, where $u(x)$ and $v(x)$ are
the particle and hole wavefunctions respectively. The single particle
Hamiltonian
\begin{equation}
H_{0}(x)=\frac{p_{x}^{2}}{2m^{\ast }}-\alpha p_{x}\sigma _{y}-V_{z}\sigma
_{z}+Z\delta (x)-\mu ,
\end{equation}%
where $m^{\ast }$ is the effective mass of the electron, $\alpha $ is the
Rashba spin-orbit coupling interaction, $V_{z}$ is the perpendicular Zeeman
field, $\mu $ is the chemical potential in the semiconductor which is
controlled by the density of doped electrons. For simplicity, we use $%
Z\delta (x)$ to model the barrier potential at $x=0$. The superconducting
order parameter $\Delta =\Delta _{0}\theta (x)$, where the step function $%
\theta (x)=1$ when $x>0$ and $\theta (x)=0$ when $x<0$.

%%%%%%%%%%%%%%%%%%%%%%%%%%%%%%%%%%%%%%%%%%%%%%%%%%%%%%%%%%%%%%%%%%
\begin{figure}[t]
\includegraphics[width = 1\linewidth]{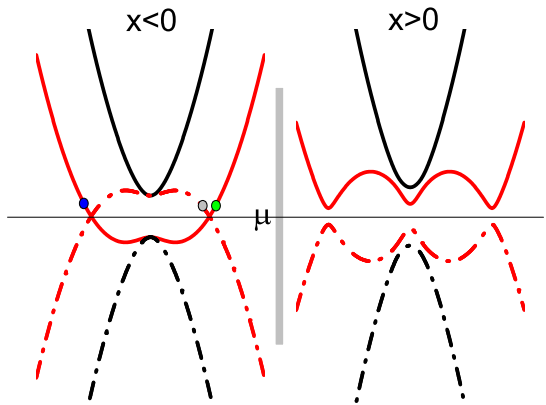}
\caption{(Color online) Energy spectrum for the semiconductor nanowire. In
the $x<0$ part, the solid (dashed dotted) lines are for electrons (holes).
Due to the superconducting coupling between hole and particle, the bulk gap
is opened in the $x>0$ part. The tunneling barrier is at $x=0$. $\protect\mu
$ is chemical potential, which is the same for both parts. Filled circles
represent incident (green), Andreev reflected (grey), and normal reflected
(blue), respectively. The dimensionless parameters are $\protect\alpha =2$, $%
\Delta _{0}=0.2$, $\protect\mu =0$, and $V_{z}=0.5$.}
\end{figure}
%%%%%%%%%%%%%%%%%%%%%%%%%%%%%%%%%%%%%%%%%%%%%%%%%%%%%%%%%%%%%%%%%%%

The energy spectrum for the left and right sections of the heterostructure
is plotted in Fig. 2. In the left section ($x<0$) with $\Delta =0$, it is
just the single particle spectrum determined by $H_{0}$, which has two bands
with the energy dispersion%
\begin{equation}
E_{\pm }=\frac{k_{x}^{2}}{2m^{\ast }}-\mu \pm \sqrt{{V_{z}}^{2}+\alpha
^{2}k_{x}^{2}}.
\end{equation}%
The energy gap at $k_{x}=0$ is $2V_{z}$. Henceforth we set $\hbar=1$. When
the chemical potential lies in the gap, electrons only occupy the lower
spin-orbit band at a low temperature. To compare with the superconducting
spectrum in the right section of the nanowire ($x>0$), the spectrum for the
hole Hamiltonian $-H_{0}^{\ast }$ is also plotted.

With the non-zero superconducting order parameter $\Delta _{0}$ in the right
section, the bulk quasiparticle excitation spectrum becomes
\begin{equation}
E^{2}=\tilde{\epsilon}^{2}+{V_{z}}^{2}+\alpha ^{2}k_{x}^{2}+\Delta
_{0}^{2}\pm 2\sqrt{V_{z}^{2}\Delta _{0}^{2}+\tilde{\epsilon}%
^{2}(V_{z}^{2}+\alpha ^{2}k_{x}^{2})},
\end{equation}%
where $\tilde{\epsilon}=\frac{k_{x}^{2}}{2m^{\ast }}-\mu $. Around the Fermi
surface, a bulk gap $2\Delta _{0}$ is always opened between the middle two
bands due to the superconducting coupling between the particle and hole
branches. However, in contrast to a regular \textit{s}-wave superconductor
where the minimum energy gap appears around the Fermi surface, the minimum
energy gap in the spin-orbit coupled system may appear at $k_{x}=0$. In
fact, the energy gap at $k_{x}=0$ disappears when
\begin{equation}
V_{z}=\sqrt{\Delta _{0}^{2}+\mu ^{2}}
\end{equation}%
\newline
although the superconducting order parameter is nonzero. When $V_{z}<\sqrt{%
\Delta _{0}^{2}+\mu ^{2}}$, the system is in a regular s-wave
non-topological state without Majorana fermions. While when $V_{z}>\sqrt{%
\Delta _{0}^{2}+\mu ^{2}}$, the system is in a topological quantum state
with Majorana fermions at $x=0$. Across $\sqrt{\Delta _{0}^{2}+\mu ^{2}}$,
the change of the Zeeman field $V_{z}$ defines a topological phase
transition from a non-topological superconductor to a topological
superconductor \cite{Sau1}. For the simplicity of the calculation, in the
following we consider three different cases: (i) $\mu =0$, and $V_{z}>\Delta
_{0}$ (topological state); (ii) $\mu =0$, and $V_{z}<\Delta _{0}$
(non-topological state with one conduction band occupied by electrons);
(iii) $\mu >$ $V_{z}$ (non-topological state with two conduction bands
occupied by electrons). We have confirmed that the results for other
parameters are similar.

\section{Electrical conductance of the nanowire}

The electrical conductance along the nanowire can be calculated based on the
BTK theory \cite{Blonder,Mizuno,Linder2}, where the wavefunctions at both
sides of the potential barrier are constructed and matched at $x=0$. We
first consider cases (i) and (ii). The wavefunction in the left section of
the nanowire can be obtained analytically as

\begin{widetext}
\begin{eqnarray}
\psi(x<0)&=&
e^{ik_+x}
\begin{pmatrix}
-i\alpha{k_+}\\
\frac{k_+^2}{2m}-V_z-E\\
0\\
0
\end{pmatrix}{\zeta_+}
+
ae^{ik_-x}
\begin{pmatrix}
0\\
0\\
i\alpha{k_-}\\
-\frac{k_-^2}{2m}+V_z-E
\end{pmatrix}{\zeta_-}
+
be^{q_-x}
\begin{pmatrix}
0\\
0\\
-\alpha{q_-}\\
-\frac{q_-^2}{2m}-V_z+E
\end{pmatrix}{\xi_+}    \\ \nonumber
&+&
ce^{-ik_+x}
\begin{pmatrix}
i\alpha{k_+}\\
\frac{k_+^2}{2m}-V_z-E\\
0\\
0
\end{pmatrix}{\zeta_+}
+
de^{q_+x}
\begin{pmatrix}
-\alpha{q_+}\\
-\frac{q_+^2}{2m}-V_z-E\\
0\\
0
\end{pmatrix}{\xi_-},
\end{eqnarray}
\end{widetext}where the wavevectors are defined as \newline
${k_{\pm }}=\sqrt{2m\left( m\alpha ^{2}\pm {E}+\sqrt{m^{2}\alpha
^{4}+V_{z}^{2}\pm 2m\alpha ^{2}E}\right) }$, ${q_{\pm }}=\sqrt{2m\left(
-m\alpha ^{2}\mp {E}+\sqrt{m^{2}\alpha ^{4}+V_{z}^{2}\pm 2m\alpha ^{2}E}%
\right) }$. $a$ and $b$ are AR coefficients for the lower and higher bands
respectively. $c$ and $d$ are the corresponding normal reflection
coefficients. $\zeta _{\pm }=1/\sqrt{\alpha ^{2}k_{\pm }^{2}+(\frac{k_{\pm
}^{2}}{2m}-V_{z}\mp {E})^{2}}$ and $\xi _{\pm }=1/\sqrt{\alpha ^{2}q_{\mp
}^{2}+(\frac{q_{\mp }^{2}}{2m}+V_{z}\mp {E})^{2}}$ are the normalization
constants of the wavefunctions. Since the chemical potential lies between
two spin-orbit bands in cases (i) and (ii), there is no plane wave solution
for the higher band, and decaying wavefunctions have been used in Eq. (6).
Because of the superconducting proximity effect, analytical expressions for
the wavefunction in the right section of the nanowire are not available.
Instead, we numerically solve the eigenvalue equations and find the
corresponding transmitted wavefunction $\psi (x>0)$ for different incident
particle energy $E$.

The wavefunctions $\psi (x)$ and its derivative $v_{x}\psi (x)$ should be
matched at the interface $x=0$ between two sections, where $v_{x}=\frac{%
\partial \mathcal{H}}{\partial p_{x}}$ is the velocity matrix determined by
the Hamiltonian (1). The reflection and transmission coefficients are
calculated by solving the resulted $8\times 8$ linear equations for the
unknown coefficients. The electrical conductance along the nanowire is%
\begin{equation}
G(E)=\left[ 1+|a(E)|^{2}-|c(E)|^{2}\right] G_{0},
\end{equation}%
with $a\left( E\right) $ ($c\left( E\right) $) is the Andreev (normal)
reflection coefficient in the lower band of $H_{0}$, $G_{0}=e^{2}/h$ is the
electrical conductance of the nanowire without the superconductor and the
magnetic field. Because the wavefunctions in the higher band are evanescent
waves when the chemical potential lies inside the gap between two spin-orbit
bands, the Andreev and normal reflection coefficients in the higher band do
not contribute to the electrical conductance and thus have been neglected.
However, the higher band wavefunctions on both sides of $x=0$ must be
included in the calculation for a self-consistent solution, where the number
of unknown coefficients should match with the number of linear equations
determined by the boundary conditions at $x=0$. In addition, we are only
interested in AR which dominates the electron transportation in the
superconducting gap region on the $x>0$ side, therefore we restrict the
incident electron energy $E$ less than the minimum quasiparticle energy gap
in our calculation.

%%%%%%%%%%%%%%%%%%%%%%%%%%%%%%%%%%%%%%%%%%%%%%%%%%%%%%%%%%%%%%%%%%
\begin{figure}[tbp]
\includegraphics[width = 1\linewidth]{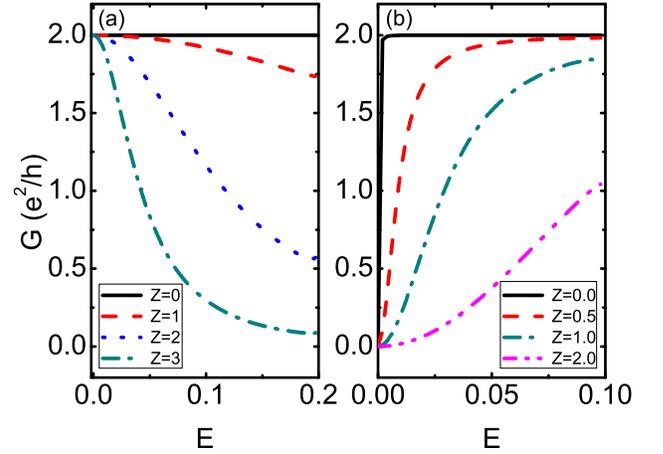}
\caption{(Color online) Plot of the electrical conductance along the
nanowire versus the incident electron energy $E$ for different barrier
heights. Other parameters are $\protect\mu =0$, $\protect\alpha =2.0$, $%
\Delta _{0}=0.2$. (a) $V_{z}=0.5$ and $V_{z}>\protect\sqrt{\Delta _{0}^{2}+%
\protect\mu ^{2}}$ for the topological state with a Majorana fermion at $x=0$%
. (b) $V_{z}=0.1$, $V_{z}<\protect\sqrt{\Delta _{0}^{2}+\protect\mu ^{2}}$
for the non-topological state without Majorana fermions.}
\end{figure}
%%%%%%%%%%%%%%%%%%%%%%%%%%%%%%%%%%%%%%%%%%%%%%%%%%%%%%%%%%%%%%%%%%%

The dependence of the electrical conductance $G\left( E\right) $ as a
function of the incident electron energy $E$ is plotted in Fig. 3 for
different parameters. In the calculation for Fig. 3b, we restrict to the
energy region $0<E<|\Delta _{0}-V_{z}|=0.1$ rather than $0<E<\Delta _{0}=0.2$
because the minimum quasiparticle excitation gap is $|\Delta _{0}-V_{z}|$
(occurs at $k_{x}=0$ when $V_{z}$ is small), instead of $\Delta _{0}$ at the
Fermi surface. In Fig. 3a, the Zeeman field $V_{z}>\sqrt{\Delta _{0}^{2}+\mu
^{2}}$ for the topological state that supports Majorana zero energy modes at
$x=0$ \cite{Sau1,Roman1}, while in Fig. 3b $V_{z}<\sqrt{\Delta _{0}^{2}+\mu
^{2}}$ for the non-topological state without Majorana fermions. All
parameters are re-scaled to dimensionless forms. The energy unit for $\mu $,
$V_{z}$, $Z$ and $\Delta _{0}$ is chosen as $\hbar ^{2}k_{0}^{2}/2m^{\ast }$%
, and the unit for $\alpha $ is $\hbar ^{2}k_{0}/2m^{\ast }$, where $k_{0}$
is a suitable chosen wavevector to match with the experimental values of $%
\Delta _{0}$. For InAs nanowire, the typical parameters are $m^{\ast }\sim
0.04m_{e}$, $\alpha \sim 0.1$ eV \AA , $V_{z}\sim 1$ K, $\Delta _{0}\sim 1$
K. As mentioned before, $\mu =0$ has been used in the calculation.

In Fig. 3, different lines correspond to different values of the barrier
height $Z$. For a regular non-topological superconductor (Fig. 3b) without
the insulating barrier ($Z=0$), the charge transport in the superconducting
gap region is dominated by the regular AR. If there is no spin-orbit
coupling, the degeneracy between spin up and down electrons is lifted by the
Zeeman field, and there is only one single channel for the AR when the
chemical potential lies in the Zeeman gap. In this case, the AR is strongly
suppressed and the electrical conductance is very small \cite{Fisher}. The
existence of spin-orbit coupling greatly enhances the AR at the finite $E$,
but the electrical conductance is still zero at $E=0$ when only a single
spin-orbit conduction band is occupied \cite{Wimmer}, as clearly seen from
Fig. 3b. For a large spin-orbit coupling, the regular AR dominates at the
finite $E$, that is, $a\left( E\right) \sim 1$ and $c\left( E\right) \sim 0$%
. Therefore $G\left( E\right) =2G_{0}$ is a constant, as observed. With the
increasing barrier height $Z$, the AR coefficient $a\left( E\right) $
decreases and normal reflection coefficient $c\left( E\right) $ increases.
For a very large $Z$, the amplitude of the AR should be zero\ (\textit{i.e.}%
, $a\left( E\right) \sim 0$, and $c\left( E\right) \sim 1$) because the
incident electrons are reflected completely by the high barrier, leading to
a vanish $G$. However, the existence of the Majorana zero energy mode at $E=0
$ in the topological state (Fig. 3a) leads to a resonant AR \cite{Law},
which gives a resonant peak at $E=0$ even for a large barrier height $Z$.
The resonant tunneling (thus the conductance) is the largest at the zero
energy, but decreases quickly at the finite energy for a large barrier
potential, as observed in Fig. 3a.

In Fig. 4, we plot the conductance peak width (defined by the energy $E_{w}$
with $G(E_{w})=G_{0}$) with respect to the barrier height $Z$ (Fig. 4a) and
spin-orbit coupling strength $\alpha $ (Fig. 4b) for the topological state
(Fig. 3a) with Majorana fermions. We see that as the barrier height
increases, the conductance peak is narrower, as expected. When the
spin-orbit coupling $\alpha $ increases, the conductance peak width is wider
and the signature of the Majorana fermions is more clear. In a practical
experiment, we should adopt a large barrier height to eliminate the regular
AR, and a large spin-orbit coupling to enhance the resonant AR induced by
the Majorana fermions.

\begin{figure}[tbp]
\includegraphics[width = 1\linewidth]{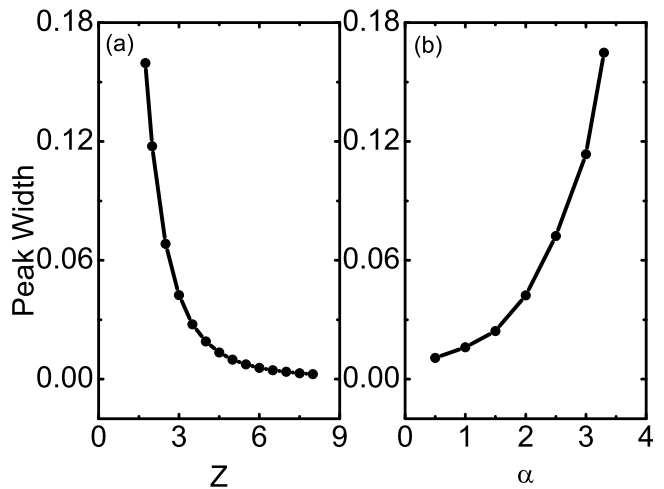}
\caption{Plot of the conductance peak width (in unit of $G_{0}=e^{2}/h$) as
a function of (a) the barrier height $Z$ and (b) the spin-orbit coupling
strength $\protect\alpha $ in the topological state (e.g., Fig. 3a). The
other parameters are $V_{z}=0.5$, $\Delta _{0}=0.2$, $\protect\mu =0.0$.}
\end{figure}

We now consider the case (iii) where the chemical potential intercepts two
conduction bands and there are no topological superconducting states. In
this case, $q_{\pm }$ in the wavefunction (6) becomes imaginary and there
exist plane wave solutions for the high spin-orbit coupled band. Following
similar procedure, we calculate the electrical conductance and plot it in
Fig. 5. In this case, the electrical conductance is $2G_{0}$ for $Z=0$ even
at $E=0$ because of the two AR channels, which is different from that in
Fig. 3b for a single AR channel. With the increasing barrier height $Z$, the
electrical conductance reduces smoothly for all $E$ in the gap region,
similar as that for a junction between a normal metal and a regular s-wave
superconductor.

\begin{figure}[b]
\includegraphics[width = 1\linewidth]{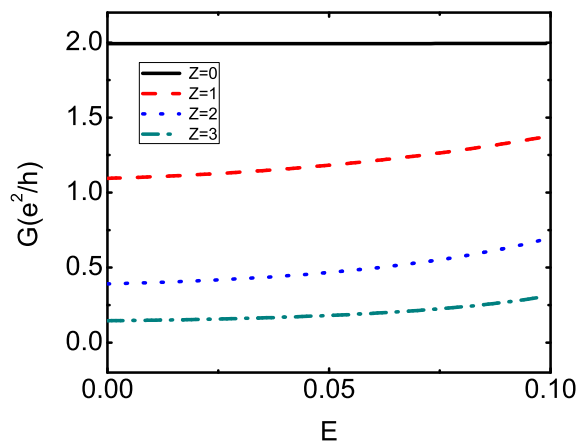}
\caption{(Color online) Plot of the electrical conductance along the
nanowire versus the incident electron energy $E$ when the chemical potential
intercepts two conduction bands. $\protect\mu =1$, $\protect\alpha =2.0$, $%
\Delta _{0}=0.2$, $V_{z}=0.5$ ($V_{z}<\protect\sqrt{\Delta _{0}^{2}+\protect%
\mu ^{2}}$ for the non-topological state).}
\end{figure}

\section{Electrical conductance in a semiconductor thin film}

In the above section, we have shown the signature of Majorana fermions in
semiconductor nanowires in a charge transport type of experiment. Note that
Majorana fermions are also proposed to exist in a two-dimensional
semiconductor thin film in proximity contact with an \textit{s}-wave
superconductor and a magnetic insulator \cite{Sau1,Mao1}. Therefore it is
natural and interesting to study whether similar transport signature can be
observed in such a 2D semiconductor heterostructure, where the single
particle Hamiltonian $H_{0}$ in Eq. (1) should be replaced with
\begin{equation}
H_{T}(x,y)=\frac{p^{2}}{2m^{\ast }}-\alpha (p_{x}\sigma _{y}-p_{y}\sigma
_{x})-V_{z}\sigma _{z}+Z\delta (x)-\mu .
\end{equation}

The energy spectrum of $H_{T}$ in the momentum space is the same as that for
the nanowire (Fig. 2) with the substitution $k\rightarrow \sqrt{%
k_{x}^{2}+k_{y}^{2}}$. In the 2D semiconductor thin film, the perpendicular
magnetic field in the nanowire (Fig. 1) is replaced with a magnetic
insulator in proximity contact with the thin film to avoid orbital effects
(such as unwanted vortices) induced by the magnetic field. On the $x>0$
side, the thin film is still in contact with an \textit{s}-wave
superconductor. In this system, the translational symmetry is broken along
the $x$ direction but conserved along the $y$ direction, therefore $k_{y}$
is a good quantum number.

The electrical conductance of the semiconductor thin film heterostructure is
calculated following the same procedure as that for the semiconductor
nanowire. The system can support Majorana zero energy states at the
interface $x=0$. The results for the normal incidence, i.e., $k_{y}=0$, are
similar as the nanowire: depending on the magnitude of the Zeeman field,
Majorana zero energy mode exists or disappears, which are characterized by
the emergence or disappearance of the conductance peak at the zero incident
energy. Therefore here we only show the results for $k_{y}\neq 0$.

\begin{figure}[tbp]
\includegraphics[width = 1\linewidth]{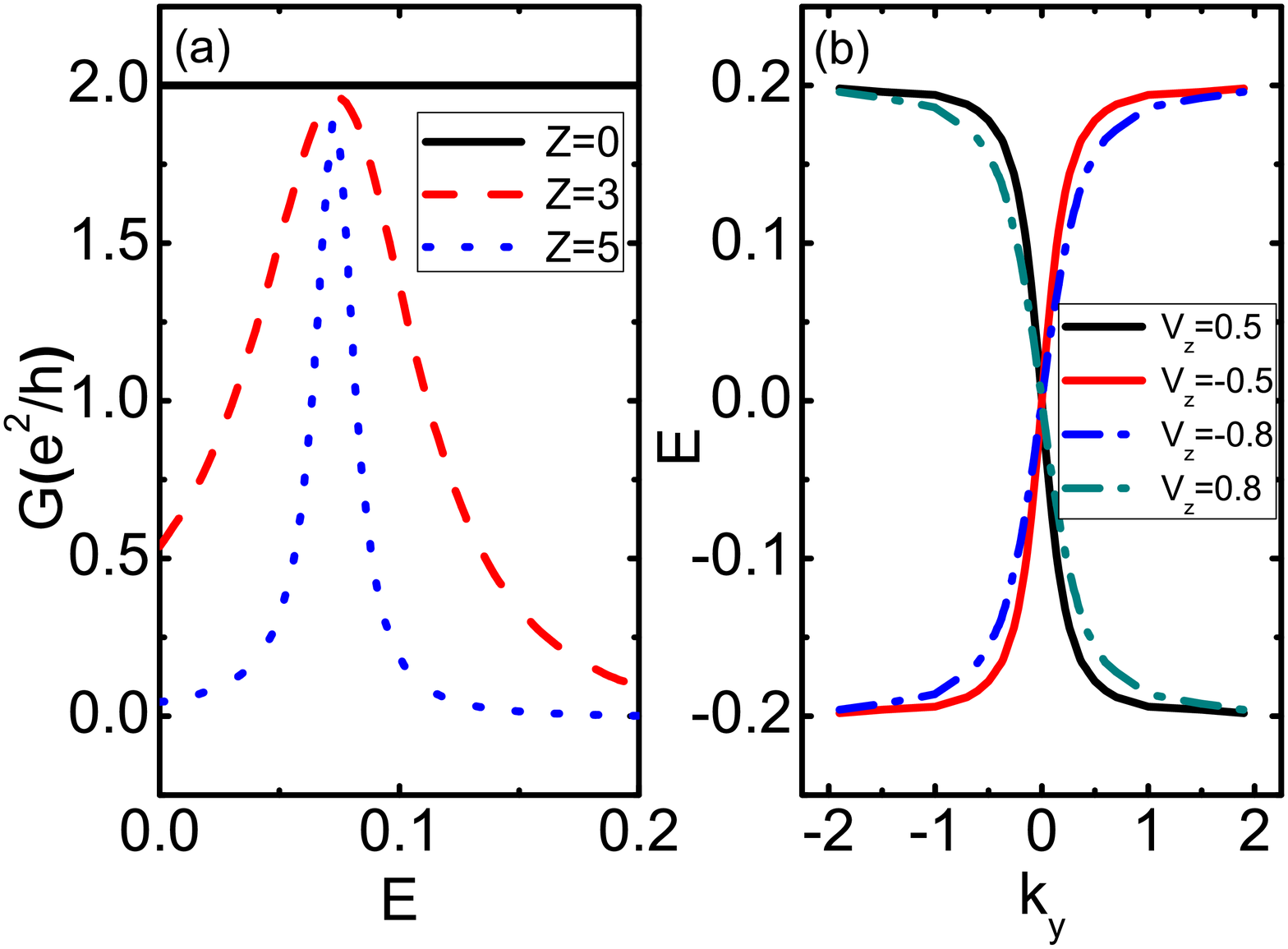}
\caption{(Color online) The electrical conductance in the 2D semiconductor
thin film. $\protect\alpha =2$, $\Delta _{0}=0.2$. (a) Plot of the
conductance with respect to the incident electron energy $E$. $k_{y}=-0.1$. $%
Z=0$ (solid line), $Z=3$ (dashed line), $Z=5$ (dotted line). (b) Edge state
dispersion relation $E\left( k_{y}\right) $. Solid lines are for $%
|V_{z}|=0.5 $ and dashed dotted lines are for $|V_{z}|=0.8$. }
\end{figure}

For certain value of $k_{y}$, we find that there is a conductance peak
existing inside the bulk gap (see Fig. 6a) in the parameter region $V_{z}>%
\sqrt{\Delta _{0}^{2}+\mu ^{2}}$. While the peaks vanishes when $V_{z}<\sqrt{%
\Delta _{0}^{2}+\mu ^{2}}$. However, the peak occurs at certain finite value
of $E$ inside the gap, which is different from the peak at $E=0$ for $%
k_{y}=0 $. The center the conductance peak moves away from $E=0$ with the
increasing $k_{y}$. This phenomena can be understood from the edge states
\cite{Hasan} at the interface $x=0$. In the 2D heterostructure at $x>0$, the
edge state at $x=0$ has an energy dispersion relation $E(k_{y})$ inside the
bulk gap between the particle and hole branches of the BdG Hamiltonian (1).
The edge state at $k_{y}$ can induce a resonant AR for electrons with energy
around $E(k_{y})$, leading to a conductance peak. In Fig. 6b, we plot the
energy dispersion for the edge state, which has the following properties:
(1) the chirality of the dispersion (\textit{i.e.}, increase or decrease
with increasing $k_{y}$) is determined by the sign of the perpendicular
Zeeman field; (2) the slope of the energy dispersion (\textit{i.e.}, the
velocity of electrons in the edge state) is determined by the magnitude of
the Zeeman field: the smaller the magnetic field, the steeper the
dispersion. We note that these two properties are the same as that of the
Andreev-bound state dispersion in the junction composed of a magnetic
insulator and an \textit{s}-wave superconductor on the surface of a
topological insulator \cite{Linder2}.

In a practical experiment, the incident electron beam may contain different $%
k_{y}$ due to the scattering to different angles from the interface,
therefore the conductance should be the summation over the signals from
different $k_{y}$. Since the edge states at other $k_{y}$ also give peaks at
non-zero $E$ through resonant AR (although the peak height does drop
slightly with the increasing $E$), it is hard to extract signals for the
Majorana zero energy states. Therefore the semiconductor nanowire would be a
better system for the observation of Majorana fermions than the
semiconductor thin film in the charge transport type of experiments.

%%%%%%%%%%%%%%%%%%%%%%%%%%%%%%%%%%%%%%%%%%%%%%%%%%%%%%%%%%%%%%%%%%

%%%%%%%%%%%%%%%%%%%%%%%%%%%%%%%%%%%%%%%%%%%%%%%%%%%%%%%%%%%%%%%%%%%

\section{Conclusion}

In summary, we study the transport properties along a semiconductor nanowire
which is partially in proximity contact with an \textit{s}-wave
superconductor and subject to a perpendicular magnetic field. We find that
Majorana fermions existing in a certain parameter region and at the
interface between the superconducting and normal parts of the nanowire can
induce resonant AR, which yields a zero energy peak in the transport
electrical conductance. We characterize the properties of the peak for
different experimental parameters. We show that such zero energy conductance
peak disappears in a 2D semiconductor thin film with the same setup because
of the existence of other edge states in 2D. We believe our proposed
transport experiment provides a simple and experimental feasible way for the
experimental detection of Majorana fermions. Finally, although the
calculation is done for electron-doped semiconductor nanowires, the Majorana
fermions induced zero energy conductance peaks should also be observable in
hole-doped semiconductor nanowires \cite{Mao3}, where Majorana fermions
exist with a higher carrier density and a lower magnetic field, and thus may
be more experimentally accessible.

\emph{Acknowledgement}--We thank M. Wimmer for helpful discussion. This work
is supported by DARPA-MTO (FA9550-10-1-0497), DARPA-YFA (N66001-10-1-4025),
and NSF-PHY (1104546).

\end{document}